\documentclass[amsmath,amssymb,aps,prd]{revtex4-1}

\usepackage[english]{babel}
\usepackage{xcolor}
\usepackage{url}
\usepackage{graphicx}
\usepackage{dcolumn}
\usepackage{bm}
\usepackage{listings} 
\usepackage{xcolor} 

\usepackage{comment}

\lstdefinelanguage{JuliaLocal}%
{%
    morekeywords={%
        abstract, baremodule, begin, break, catch, const, continue, do, else, elseif, end, export, false, finally, for, function, global, if, import, importall, in, isa, let, local, macro, module, mutable, new, primitive, quote, return, struct, true, try, type, typealias, using, while, where,
        abs, abs2, abspath, accept, accumulate, acos, acosh, all, all!, any, append!, append, apply, asin, asinh, atan, atanh, broadcast, cat, cd, ceil, chdir, chmod, chomp, chop, circshift, clear!, close, collect, conj, connect, contains, continue, copy, cos, cosh, count, countlines, cross, csc, csch, ctime, cumprod, cumsum, det, diagm, diff, div, dot, eachindex, eigen, eigvals, eigmax, eigmin, eigvecs, endswith, eval, exp, expm, expm1, fill, filter, find, findall, findfirst, findlast, findmax, findmin, findnext, findprev, first, fld, fld1, floor, flush, foldl, foldr, foreach, frexp, fullname, get, get!, getfield, gethostname, getpid, getproperty, getsockname, haskey, hasproperty, hcat, hex, hms, host2ip, hypot, identity, imag, indexin, indices, insert!, inv, invmod, isapprox, isascii, isdigit, isempty, isequal, isfinite, isinf, isnan, isopen, ispath, ispow2, isqrt, isreadable, isready, issubnormal, istaskdone, iterate, join, keys, kron, ldexp, lcm, length, linreg, log, log10, log1p, log2, logb, lower, lpad, lstat, lstrip, map, map!, mapfoldl, mapfoldr, match, max, maximum, maximum!, mean, median, merge, merge!, min, minimum, minimum!, mkdir, mod, mod1, modf, mv, nextafter, nextfloat, norm, normpath, not, ntuple, one, ones, open, or, parse, path, permute!, pinv, pop!, prepend!, prevfloat, prod, push!, pwd, rand, randn, range, read, read!, readdir, readline, readlines, reduce, reduce1, rem, repeat, replace, reset, reshape, resize!, reverse, reverse!, rm, round, rpad, rsearch, rstrip, run, schedule, searchsorted, searchsortedfirst, searchsortedlast, sec, sech, setdiff, setdiff!, setprecision, shift!, shuffle, shuffle!, sign, signbit, sin, sinh, size, sleep, sort, sort!, sorted, splice!, split, splitdrive, splitext, splitpath, sqrt, square, start, step, steps, stop, str, strip, sub2ind, sum, sum!, svd, svdvals, symdiff, symdiff!, take!, tan, tanh, tempname, text, thisind, tic, time, toc, toq, touch, tr, trues, trunc, trunc, trylock, typeassert, typejoin, typeintersect, typeof, typemax, typemin, unescape, union, union!, unique, unique!, unlock, unshift!, uppercase, vcat, vec, view, wait, walkdir, which, while, widemul, with, write, write!, xor, yield, yieldto, zeros, zip,
        NaN, Inf, nothing
    },
    sensitive=true,
    morecomment=[l]{\#},
    morecomment=[n]{\#=}{=\#},
    morestring=[s]{"}{"},
    morestring=[m]{'}{'},
}

\lstdefinestyle{julia}{
    backgroundcolor=\color[HTML]{F2F2F2},   
    basicstyle=\ttfamily\small\color[HTML]{19177C}, 
    keywordstyle=\color[HTML]{1BA1EA},      
    stringstyle=\color[HTML]{F5615C},       
    commentstyle=\color[HTML]{AAAAAA},      
    numberstyle=\tiny\color[HTML]{7F7F7F},  
    rulecolor=\color{black},                
    frame=tb,                               
    framesep=5pt,                           
    framexleftmargin=5pt,                   
    framexrightmargin=5pt,                  
    tabsize=4,                              
    captionpos=b,                           
    breaklines=true,                        
    breakatwhitespace=false,                
    showstringspaces=false,                 
    showspaces=false,                       
    showtabs=false,                         
    columns=fullflexible,                   
    keepspaces=true,                        
    numbers=left,                           
    stepnumber=1,                           
}

\usepackage{hyperref}
\hypersetup{
    colorlinks=true,
    linkcolor=blue,
    filecolor=magenta,      
    urlcolor=cyan,
} 
\begin{document}

\title{JuliaQCD: Portable lattice QCD package in Julia language}

\author{Yuki Nagai}
\email{nagai.yuki@mail.u-tokyo.ac.jp}
\affiliation{Information Technology Center, The University of Tokyo, 6-2-3 Kashiwanoha, Kashiwa, Chiba 277-0882, Japan}

\author{Akio Tomiya}%
\email{akio@yukawa.kyoto-u.ac.jp}
\affiliation{
Department of Information and Mathematical Sciences, Tokyo Woman’s Christian University, Tokyo 167-8585, Japan
}
\affiliation{
RIKEN Center for Computational Science, Kobe 650-0047, Japan
}

\date{\today}

\begin{abstract}
We present JuliaQCD, a new lattice gauge theory codebase developed in the Julia language. Julia is well suited to integrating machine-learning techniques and enables rapid prototyping and execution of algorithms for four-dimensional QCD and other non-Abelian gauge theories. The code leverages LLVM for high-performance execution and supports MPI for parallel computing. Julia's multiple dispatch provides a flexible and intuitive framework for development. The code implements existing algorithms such as Hybrid Monte Carlo (HMC) and supports many colors and flavors, lattice fermions, smearing techniques, and full QCD simulations. It is designed to run efficiently across various platforms, from laptops to supercomputers, allowing seamless scalability. The codebase is currently available on GitHub at \url{https://github.com/JuliaQCD}.
\end{abstract}

\maketitle

\section{Introduction}
Quantum Chromodynamics (QCD) is the fundamental theory describing the interactions of quarks and gluons, the elementary particles that make up the subatomic world. 
Because they can be created and annihilated from the vacuum by relativistic processes, they must be treated as a quantum many-body system with an indeterminate particle number. 
QCD is a quantum field theory that accurately describes our universe. 

However, perturbation theory fails in QCD due to the strong coupling. Lattice QCD addresses this challenge by employing lattice regularization, introducing a discrete spacetime cutoff to regulate the path integral. This approach enables the computation of quantum expectation values of physical observables.

Lattice QCD was first formulated by K. G. Wilson in 1974 \cite{Wilson:1974sk}, and numerical studies of QCD using this method were initiated by M. Creutz \cite{Creutz:1980zw}. This framework enables detailed studies of the QCD vacuum structure and provides precise insights into the Standard Model. Additionally, it facilitates investigations into phenomena such as the quark-gluon plasma and dark-matter candidates.

To formulate lattice QCD, a spacetime lattice cutoff is introduced. This cutoff is not merely a numerical approximation; it fundamentally defines QCD itself. Traditionally, quantum field theory is formulated on a continuous spacetime, but in the presence of interactions such theories suffer from divergences. Although these divergences can be renormalized, they are inherent in the continuum formulation.

In QCD, we aim to calculate the quantum expectation value of a gauge-invariant operator $O$ \footnote{For certain purposes, we sometimes calculate the expectation value of a gauge-{\it variant} operator. In such cases, gauge fixing is required by Elitzur's theorem \cite{Elitzur:1975im}.}. In the path integral formalism, this is expressed as
\begin{align}
\langle O \rangle
=
\frac{1}{Z}
\int \mathcal{D}U \mathcal{D}\bar{q} \mathcal{D}q\; \mathrm{e}^{-S_\text{g}[U] - S_\text{f}[U, q, \bar{q}]} O(U, q, \bar{q}),
\end{align}
where $q$ and $\bar{q}$ are the quark and antiquark fields, respectively, and $U$ is the gauge field on the lattice. The term $S_\text{g}[U]$ represents the gauge field action, and $S_\text{f}[U, q, \bar{q}]$ denotes the action for the quarks. The factor $Z$ is a normalization constant ensuring that $\langle 1 \rangle = 1$. 
One can obtain observables in QCD by evaluating this integral. 

The standard algorithm used for simulating lattice QCD is the Hybrid Monte Carlo (HMC) algorithm \cite{Duane:1987de}. HMC has become the de facto standard because of its efficiency and effectiveness in dealing with the high-dimensional integrals for lattice gauge fields with dynamical fermions. It has been successfully used in a wide range of studies, from calculating the hadron spectrum \cite{CP-PACS:2001vqx} and form factors to determining the critical temperature ($T_c$) of QCD. The success of lattice QCD in these areas highlights the power of numerical methods in exploring nonperturbative aspects of quantum field theory.

Recent developments in machine learning (ML) have opened new avenues for enhancing lattice QCD calculations. Many researchers are developing new algorithms that integrate ML techniques into lattice QCD simulations. Prototyping is crucial in this context, as we aim to create efficient algorithms for lattice QCD that leverage ML techniques. One of the central trade-offs in this process is between code simplicity and execution speed. While ease of implementation is important, code that is too slow is unusable in practice. Therefore, we focus on optimizing the total time required for lattice QCD studies, which includes both the time spent developing the algorithm and the time required for its execution.

Portability and parallelizability are also crucial considerations. The code must run efficiently on a wide range of platforms, from laptops to supercomputers, and should be optimized to reduce the overall computational time. This allows researchers to take advantage of the available computational resources and scale their simulations as needed.

The Julia language is a new open-source scientific programming language built on LLVM \cite{bezanson2012juliafastdynamiclanguage}. 
Unlike interpreters, Julia is a compiled language with a just-in-time (JIT) compiler. This allows it to achieve performance comparable to Fortran and C \cite{julia_benchmarks}, while maintaining the simplicity of Python. 
Julia's flexibility and speed make it an ideal choice for developing new lattice QCD algorithms, as it allows for rapid prototyping without sacrificing performance.
Thanks to these advantages, Julia is an attractive language for high-energy physics \cite{Eschle:2023ikn}.

In this paper we introduce a new lattice QCD code, JuliaQCD, written in the Julia language.
The motivation behind the development of this codebase is to minimize the combined time of algorithm implementation and execution. Existing languages such as Python are too slow for large-scale lattice QCD simulations, while C++/C/Fortran present challenges in terms of ease of use and development time. Julia strikes the right balance, offering both the speed required for intensive computations and the simplicity needed for fast prototyping. Our goal is to create a flexible and high-performance framework that can be easily adapted to new developments in both lattice QCD and machine learning.

\subsection{JuliaQCD project}
JuliaQCD is the first open-source lattice QCD code suite in the Julia programming language. 
(Ultimate) high performance is not within our scope, but we focus on prototyping with good scalability.
We developed this code with the following goals:
\begin{enumerate}
    \item Excellent portability 
    \item Quick and easy setup
    \item Comprehensive suite 
    \item Highly modifiable
    \item Competitive performance 
\end{enumerate}
The system offers excellent portability, as it runs smoothly on any machine with Julia installed, without the need for complicated setup processes. It is also quick and easy to set up, enabling users to start working in less than 5 minutes without requiring complex compilation steps. In addition, the system provides a comprehensive suite that supports both full QCD and quenched configuration generation, along with a variety of measurement tools. Its highly modifiable structure makes it ideal for rapid prototyping and experimenting with different approaches. Finally, the system delivers competitive performance, comparable to traditional Fortran 90 codes, ensuring efficiency without sacrificing flexibility.

\subsection{Related Work}
In the broader landscape, traditional codes such as the MILC code and the Columbia Physics System (CPS) have also played significant roles. Despite their long history, these codes remain functional and are used in various research contexts. MILC, in particular, has been a cornerstone for many researchers, supporting a wide range of computations, including those utilizing GPU architectures \cite{milccode}. CPS continues to be a relevant reference point for its C-based implementation.

FermiQCD \cite{DiPierro:2001yu} has also contributed to the field. FermiQCD, a collection of C++ classes and parallel algorithms, facilitates fast development of lattice applications and includes optimizations for cluster computing, such as SSE2 instructions. Its focus on distributed processing has allowed it to remain an important tool in parallel lattice QCD.

OpenQCD has established itself as a flexible and advanced package for lattice QCD simulations, particularly in high-performance environments. It supports a wide range of theories, including O(a)-improved Wilson quarks, and includes specialized algorithms such as HMC and SMD. OpenQCD's flexibility allows for large-scale simulations and efficient use of contemporary processors, making it a widely used tool for complex lattice QCD computations.

The development of lattice QCD codes has been significantly influenced by contributions from various international communities, including notable efforts from Japan. Two key projects, the Lattice Tool Kit in Fortran 90 (LTK) \cite{Choe:2002pu} and Bridge++ \cite{Akahoshi:2021gvk}, originate from this community. LTK, primarily written in Fortran, has served as a foundational tool, though it is not optimized for contemporary high-performance computing (HPC) environments. In contrast, Bridge++ was developed with a focus on providing a comprehensive and efficient toolset for lattice QCD simulations on modern supercomputers. Both projects reflect Japan's strong engagement in advancing computational techniques for lattice QCD.

In recent years, there has been a trend toward utilizing C++, as in Bridge++, for its advanced features and improved performance in complex simulations. Grid \cite{Boyle:2015tjk} is a prime example of this trend, offering robust support for both CPU and GPU computations. The Grid Python Toolkit (GPT) extends this functionality to Python, allowing for easy integration with data analysis and machine learning workflows. Another significant development is SIMULATeQCD \cite{HotQCD:2023ghu}, which specializes in multi-GPU implementations, enabling high-performance lattice QCD computations. Additionally, QUDA (QCD on CUDA) \cite{Clark:2009wm} serves as a sub-library that provides specialized routines for running lattice QCD computations on NVIDIA GPUs. By leveraging CUDA architectures, QUDA offers substantial speedups for specific operations within larger QCD packages, rather than functioning as a standalone QCD software suite.

The increasing adoption of Python in scientific computing has led to the development of tools such as Lyncs-API, pyQCD \cite{Spraggs:2014pda}, and the Grid Python Toolkit (GPT). These projects provide high-level interfaces for lattice QCD calculations, combining Python's user-friendly syntax with the computational efficiency of C++ backends. Lyncs-API, as discussed in \cite{Bacchio:2022bjk}, offers a flexible API for a variety of QCD computations, while pyQCD similarly integrates Python and C++ for efficient lattice QCD research.
This landscape illustrates the evolution of lattice QCD software from traditional Fortran and C implementations to modern, multi-language frameworks designed for contemporary HPC environments. LatticeQCD.jl contributes to this ongoing evolution by leveraging Julia's capabilities, providing a new, efficient, and accessible option for researchers.
\section{Software description}
We support lattice gauge theory in 4-dimensional Euclidean spacetime.
We cover most standard techniques.
We also implement a wizard for parameter files.
Our code, LatticeQCD.jl, has the following functionalities.

\vspace{5mm}
\begin{minipage}{0.44\textwidth}
{\bf Gauge fields}
\begin{itemize}
\item Optimized $\mathrm{SU}(2)$, $\mathrm{SU}(3)$
\item General $\mathrm{SU}(N_c)$
\end{itemize}
{\bf Fermion actions}
\begin{itemize}
\item Wilson (2 flavors)
\item Staggered fermions (1-8 tastes)
\item Standard domain-wall (2 flavors)
\end{itemize}
{\bf Measurements}
    \begin{itemize}
        \item Plaquette
        \item Polyakov loop
        \item Chiral condensates (Wilson, staggered)
        \item Momentum-projected pion correlator (Wilson, staggered)
        \item $R\times T$ Wilson loop
        \item Energy density
        \item Topological charge (plaquette, clover, and $\mathcal{O}(a^2)$-improved definition)
        \item Load \& measurement mode (load and measure all configurations in a directory)
    \end{itemize}
    {\bf Smearing}
    \begin{itemize}
        \item Stout
        \item Gradient flow for a generic action 
    \end{itemize}
\end{minipage}
\begin{minipage}{0.44\textwidth}
{\bf Configuration generation algorithms}
\begin{itemize}
\item Cold/Hot start for $\mathrm{SU}(N_c)$. One instanton configuration for $\mathrm{SU}(2)$
\item Heatbath for $\mathrm{SU}(N_c)$ \& overrelaxation for a general gauge action
\item Even-odd heatbath for the plaquette action
\item Quenched HMC with $\mathrm{SU}(N_c)$ for a general gauge action
\item HMC (2-flavor Wilson) for $\mathrm{SU}(N_c)$ with a general gauge action
\item HMC (4-taste staggered fermions) for $\mathrm{SU}(N_c)$ with a general gauge action
\item RHMC (any-flavor staggered) for $\mathrm{SU}(N_c)$ with a general gauge action
\item Stout-smeared dynamical fermions for $\mathrm{SU}(N_c)$ 
\item Self-learning HMC with the plaquette action
\end{itemize}
{\bf I/O for gauge configurations}
    \begin{itemize}
        \item ILDG format (binary)
        \item JLD format (Julia's default binary file, HDF5-based)
        \item Bridge++ text file (Bridgetext)
    \end{itemize}
\end{minipage}

\vspace{5mm}
If one specifies other than $N_f=4, 8$ with the staggered-fermion HMC, RHMC is automatically used. For a machine with Apple Silicon, $N_f=1-8$ is available.
Parallelization is supported by MPI, which can be used with 
\href{https://github.com/akio-tomiya/Gaugefields.jl}{Gaugefields.jl} and 
\href{https://github.com/akio-tomiya/LatticeDiracOperators.jl}{LatticeDiracOperators.jl} (see Appendix \ref{sec HMC with MPI}).
We perform consistency checks against several references. See
Appendix \ref{sec:Consistency check}.

\subsection{Code Structure}
JuliaQCD is the project name, and it is composed of several packages.
\begin{itemize}
    \item \href{https://github.com/akio-tomiya/LatticeQCD.jl}{LatticeQCD.jl} : LatticeQCD.jl is a wrapper of the following packages. The interface of this package is user-friendly.
    \item \href{https://github.com/akio-tomiya/Wilsonloop.jl}{Wilsonloop.jl}: Wilsonloop.jl helps handle Wilson loops and generic Wilson lines in any $N_c$ and dimensions. Wilson lines can be defined symbolically.
    \item \href{https://github.com/akio-tomiya/Gaugefields.jl}{Gaugefields.jl}: Gaugefields.jl is a package for lattice $\mathrm{SU}(N_c)$ gauge fields. It handles gauge fields (links), gauge actions with MPI, and autograd. It can generate quenched configurations.
    \item \href{https://github.com/akio-tomiya/LatticeDiracOperators.jl}{LatticeDiracOperators.jl}: LatticeDiracOperators.jl is a package for Dirac operators and fermions on the lattice. It handles pseudofermion fields with various lattice Dirac operators, and fermion actions with MPI. It can generate configurations with dynamical fermions.
    \item \href{https://github.com/akio-tomiya/QCDMeasurements.jl}{QCDMeasurements.jl}: QCDMeasurements.jl is a package for measuring physical quantities. It includes measurements for basic quantities like chiral condensates, plaquettes, pion correlators, and topological charge with several definitions. It also includes the gradient flow with several actions.
\end{itemize}
\begin{figure}[h]
\begin{center}
\includegraphics[scale=0.5]{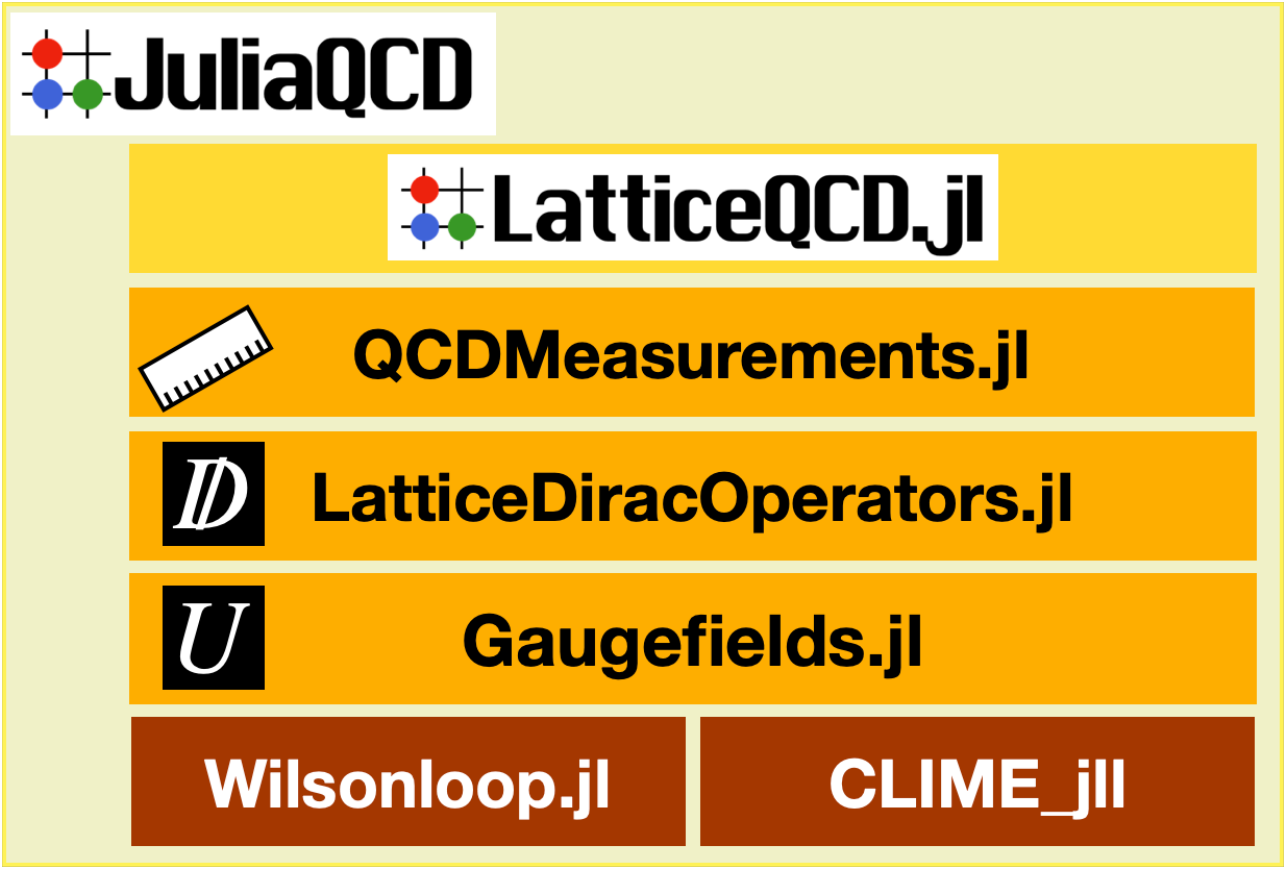}
\end{center}
  \caption{Structure of the packages (JuliaQCD).
   \label{fig:structure_package}}
\end{figure}


\section{Usage of LatticeQCD.jl}
We present example code to perform lattice QCD calculations using LatticeQCD.jl.
One can start a lattice QCD calculation in 5 steps.
\begin{enumerate}
    \item Install the Julia language from \href{https://julialang.org/downloads/}{JuliaLang.org}.  
    \item In Julia REPL, press the \texttt{]} key to enter the package mode and type:
\begin{lstlisting}[language=JuliaLocal,style=julia]
add LatticeQCD
\end{lstlisting}
    then press the ``return'' key. Press the ``backspace'' key (or ``delete'' key on Mac) to exit package mode. The latest version is available via \texttt{add LatticeQCD\#master}. All dependencies will be resolved automatically.
    \item Include the package with:
\begin{lstlisting}[language=JuliaLocal,style=julia]
using LatticeQCD
\end{lstlisting}
    \item Make a parameter file with the wizard:
\begin{lstlisting}[language=JuliaLocal,style=julia]
run_wizard()
\end{lstlisting}
    Choose parameters.
    \item Start the simulation with the created parameter file:
\begin{lstlisting}[language=JuliaLocal,style=julia]
run_LQCD("my_parameters.toml")
\end{lstlisting}
One will obtain results.
One can write a parameter file.
\end{enumerate}

\subsection{User interfaces}
We support the following two user interfaces for LatticeQCD.jl:
\begin{enumerate}
    \item Julia REPL interface (for beginners, just after the lattice QCD textbook)
    \item General interface (experience with another code, for batch jobs, customized purposes)
\end{enumerate}
Usage 1 is explained above. 

For Usage 2, in Julia REPL, press the \texttt{]} key to enter the package mode and type:
\begin{lstlisting}[language=JuliaLocal,style=julia]
add LatticeQCD
\end{lstlisting}
Then, LatticeQCD.jl will be installed.
The ``PARAMETER\_FILE'' can be created through the wizard. To use the wizard in the shell, please write the following code and save it as \texttt{wizard.jl}:
\begin{lstlisting}[language=JuliaLocal,style=julia]
using LatticeQCD
run_wizard()
\end{lstlisting}
Then, one can run the wizard:
\begin{lstlisting}[language=JuliaLocal,style=julia]
julia wizard.jl
\end{lstlisting}
Please write the following code and save it as \texttt{run.jl}:
\begin{lstlisting}[language=JuliaLocal,style=julia]
using LatticeQCD
run_LQCD(ARGS[1])
\end{lstlisting}
Then, one can execute it like this:
\begin{lstlisting}[language=JuliaLocal,style=julia]
julia run.jl PARAMETER_FILE
\end{lstlisting}

\section{Conclusion and Outlook}

In this work, we introduce JuliaQCD, a new and flexible lattice QCD codebase developed in the Julia programming language. By leveraging Julia’s high-performance just-in-time compilation and ease of use, this project aims to strike a balance between rapid prototyping and computational efficiency. JuliaQCD offers a range of functionalities, including support for full QCD simulations and various smearing techniques, while maintaining scalability across different computational environments from personal systems to high-performance computing clusters.

Our approach is motivated by the need to reduce the time required for both algorithm implementation and execution, allowing researchers to experiment with and refine techniques without sacrificing performance. The portability and ease of setup provided by JuliaQCD make it an ideal tool for researchers working at the intersection of lattice QCD and machine learning, where quick adaptation to new algorithms and methodologies is critical.

While the performance of JuliaQCD is comparable to established Fortran codes, its focus remains on flexibility and usability, providing a framework that can be easily extended to incorporate modern computational techniques. Future work will explore further optimizations, including deeper integration with GPU architectures and advanced machine learning techniques to further enhance the capabilities of lattice QCD simulations.

JuliaQCD represents an important step in the evolution of lattice QCD software, contributing to the broader trend of adopting more modern, user-friendly languages without compromising on performance. We anticipate that this work will help advance both the development and application of lattice QCD, facilitating new discoveries in Quantum Chromodynamics and related fields.


\begin{acknowledgments}
The authors thank to the \href{https://github.com/tsuchim/Lattice-Tool-Kit}{Lattice Tool Kit} written in Fortran 90.
The authors thank to Taku Izubuchi, Issaku Kanamori, Okuto Morikawa, Satoshi Terasaki and Hiromasa Watanabe for comments and contributions. 
The work of authors was partially supported by JSPS KAKENHI Grant Numbers 20K14479,
22K03539, 22H05112, and 22H05111, and MEXT as ``Program for Promoting Researches on the Supercomputer
Fugaku'' (Simulation for basic science: approaching the new quantum era; Grant Number JPMXP1020230411, and
Search for physics beyond the standard model using large-scale lattice QCD simulation and development of AI
technology toward next-generation lattice QCD; Grant Number JPMXP1020230409).

\end{acknowledgments}

\bibliography{juliaqcd}
\clearpage

\appendix
\section{Wilsonloop.jl}
A gauge action is constructed by gauge invariant objects, Wilson loops, in discretized spacetime. Wilsonloop.jl helps us to treat with the Wilson loops and generic Wilson lines in any Nc and dimensions.
Wilsonloop.jl has the following functionalities: 
\begin{itemize}
    \item From a symbolic definition of Wilson lines, this returns $\mathrm{SU}(N_c)$-valued Wilson lines as objects
    \item Constructing all staples from given symbolic Wilson lines
    \item Constructing derivatives of given symbolic Wilson lines (auto-grad for $\mathrm{SU}(N_c)$ variables)
\end{itemize}
We can easily generate a plaquette shown as 
\begin{lstlisting}[language=JuliaLocal,style=julia]
println("plaq")
plaq = make_plaq()
display(plaq)
\end{lstlisting}
The staple of the plaquette is given as 
\begin{lstlisting}[language=JuliaLocal,style=julia]
for μ=1:4
    println("μ = $μ")
    staples = make_plaq_staple(μ)
    display(staples)
end
\end{lstlisting}
An arbitrary Wilson loop is constructed as 
\begin{lstlisting}[language=JuliaLocal,style=julia]
loop = [(1,+1),(2,+1),(1,-1),(2,-1)]
println(loop)
w = Wilsonline(loop)
println("P: ")
show(w)
\end{lstlisting}
Its adjoint is calculated as 
\begin{lstlisting}[language=JuliaLocal,style=julia]
println("P^+: ")
show(w')
\end{lstlisting}
The derivative of the lines $dw/dU_{\mu}$ is calculated as 
\begin{lstlisting}[language=JuliaLocal,style=julia]
for μ=1:4
    dU = derive_U(w,μ)
    for i=1:length(dU)
        show(dU[i])
    end
end
\end{lstlisting}
Note that the derivative is a rank-4 tensor.
The derivatives are usually used for making the smearing of the gauge fields (Stout smearing can be used in Gaugefields.jl).

\section{Gaugefields.jl}

This package has following functionalities
\begin{itemize}
\item \textbf{$\mathrm{SU}(N_c)$ (Nc > 1) gauge fields in 2 or 4 dimensions with arbitrary actions.}
\item \textbf{$Z(N_c)$ 2-form gauge fields in 4 dimensions, which are given as 't Hooft flux.}
\item \textbf{U(1) gauge fields in 2 dimensions with arbitrary actions.}
\item \textbf{Configuration generation}
\begin{itemize}
    \item Heatbath
    \item quenched Hybrid Monte Carlo
    \item quenched Hybrid Monte Carlo being subject to 't Hooft twisted b.c.
with external (non-dynamical) $Z(N_c)$ 2-form gauge fields \footnote{Thanks for O. Morikawa.}
    \item quenched Hybrid Monte Carlo for $\mathrm{SU}(N_c)$/$Z(N_c)$ gauge theory
with dynamical $Z(N_c)$ 2-form gauge fields
\end{itemize}
\item \textbf{Gradient flow via RK3}
\begin{itemize}
    \item Yang-Mills gradient flow
    \item Yang-Mills gradient flow being subject to 't Hooft twisted b.c.
    \item Gradient flow for $\mathrm{SU}(N_c)$/$Z(N_c)$ gauge theory
\end{itemize}
\item \textbf{I/O: ILDG and Bridge++ formats are supported }(c-lime will be installed implicitly with CLIME\_jll )

\item \textbf{MPI parallel computation}(experimental)

\begin{itemize}
    \item quenched HMC with MPI being subject to 't Hooft twisted b.c.
\end{itemize}
\end{itemize}

In addition, this supports followings
\begin{itemize}
    \item Autograd for functions with $\mathrm{SU}(N_c)$ variables
    \item Stout smearing
    \item Stout force via backpropagation
\end{itemize}
We note that Autograd can be worked for general Wilson lines except for ones have overlaps. 

\subsection{File loading and saving}
\subsubsection{ILDG format}
We can use ILDG format, one of standard formats for configurations. 
We can read ILDG format as follows:
\begin{lstlisting}[language=JuliaLocal,style=julia]
using Gaugefields

NX = 4
NY = 4
NZ = 4
NT = 4
NC = 3
Nwing = 1
Dim = 4
U = Initialize_Gaugefields(NC,Nwing,NX,NY,NZ,NT,condition = "cold")
filename = "hoge.ildg"
ildg = ILDG(filename)
i = 1
L = [NX,NY,NZ,NT]
load_gaugefield!(U,i,ildg,L,NC)
\end{lstlisting}
With the use of the configuration, we can calculate the plaquette: 
\begin{lstlisting}[language=JuliaLocal,style=julia]
temp1 = similar(U[1])
temp2 = similar(U[1])

comb = 6
factor = 1/(comb*U[1].NV*U[1].NC)
@time plaq_t = calculate_Plaquette(U,temp1,temp2)*factor
println("plaq_t = $plaq_t")
poly = calculate_Polyakov_loop(U,temp1,temp2) 
println("polyakov loop = $(real(poly)) $(imag(poly))")
\end{lstlisting}
We can write a configuration as the ILDG format as follows: 
\begin{lstlisting}[language=JuliaLocal,style=julia]
filename = "hoge.ildg"
save_binarydata(U,filename)
\end{lstlisting}
\subsubsection{Text format for Bridge++}
Gaugefields.jl also supports a text format for Bridge++. 
A file loading and saving are expressed as follows: 
\begin{lstlisting}[language=JuliaLocal,style=julia]
using Gaugefields
filename = "testconf.txt"
load_BridgeText!(filename,U,L,NC)
\end{lstlisting}
\begin{lstlisting}[language=JuliaLocal,style=julia]
filename = "testconf.txt"
save_textdata(U,filename)
\end{lstlisting}

\subsection{Gradient flow}
To smear Gauge fields is important in LatticeQCD. 
We show the codes of the Lüscher's gradient flow as follows.
\begin{lstlisting}[language=JuliaLocal,style=julia]
NX = 4
NY = 4
NZ = 4
NT = 4
Nwing = 0
NC = 3

U = Initialize_Gaugefields(NC, Nwing, NX, NY, NZ, NT, condition="hot")

temp1 = similar(U[1])
temp2 = similar(U[1])
comb = 6
factor = 1 / (comb * U[1].NV * U[1].NC)

g = Gradientflow(U)
for itrj = 1:100
    flow!(U, g)
    @time plaq_t = calculate_Plaquette(U, temp1, temp2) * factor
    println("$itrj plaq_t = $plaq_t")
    poly = calculate_Polyakov_loop(U, temp1, temp2)
    println("$itrj polyakov loop = $(real(poly)) $(imag(poly))")
end
\end{lstlisting}

\subsection{Hybrid Monte Carlo}
With the use of the Gaugefields.jl, we can easily make the Hybrid Monte Carlo method as following. 
\begin{lstlisting}[language=JuliaLocal,style=julia]
using Random
using Gaugefields
using LinearAlgebra

function calc_action(gauge_action, U, p)
    NC = U[1].NC
    Sg = -evaluate_GaugeAction(gauge_action, U) / NC 
    Sp = p * p / 2
    S = Sp + Sg
    return real(S)
end

function MDstep!(gauge_action, U, p, MDsteps, Dim, Uold, temp1, temp2)
    Δτ = 1.0 / MDsteps
    gauss_distribution!(p)
    Sold = calc_action(gauge_action, U, p)
    substitute_U!(Uold, U)

    for itrj = 1:MDsteps
        U_update!(U, p, 0.5, Δτ, Dim, gauge_action)

        P_update!(U, p, 1.0, Δτ, Dim, gauge_action, temp1, temp2)

        U_update!(U, p, 0.5, Δτ, Dim, gauge_action)
    end
    Snew = calc_action(gauge_action, U, p)
    println("Sold = $Sold, Snew = $Snew")
    println("Snew - Sold = $(Snew-Sold)")
    ratio = min(1, exp(-Snew + Sold))
    if rand() > ratio
        substitute_U!(U, Uold)
        return false
    else
        return true
    end
end

function U_update!(U, p, ϵ, Δτ, Dim, gauge_action)
    temps = get_temporary_gaugefields(gauge_action)
    temp1 = temps[1]
    temp2 = temps[2]
    expU = temps[3]
    W = temps[4]
    for μ = 1:Dim
        exptU!(expU, ϵ * Δτ, p[μ], [temp1, temp2])
        mul!(W, expU, U[μ])
        substitute_U!(U[μ], W)
    end
end

function P_update!(U, p, ϵ, Δτ, Dim, gauge_action, temp1, temp2) # p -> p +factor*U*dSdUμ
    NC = U[1].NC
    temp = temp1
    dSdUμ = temp2
    factor = -ϵ * Δτ / (NC)
    for μ = 1:Dim
        calc_dSdUμ!(dSdUμ, gauge_action, μ, U)
        mul!(temp, U[μ], dSdUμ) # U*dSdUμ
        Traceless_antihermitian_add!(p[μ], factor, temp)
    end
end
function HMC_test_4D(NX, NY, NZ, NT, NC, β)
    Dim = 4
    Nwing = 0
    Random.seed!(123)
    U = Initialize_Gaugefields(NC, Nwing, NX, NY, NZ, NT, condition="hot", randomnumber="Reproducible")
    println(typeof(U))
    temp1 = similar(U[1])
    temp2 = similar(U[1])
    comb = 6 #4*3/2
    factor = 1 / (comb * U[1].NV * U[1].NC)
    @time plaq_t = calculate_Plaquette(U, temp1, temp2) * factor
    println("0 plaq_t = $plaq_t")
    poly = calculate_Polyakov_loop(U, temp1, temp2)
    println("0 polyakov loop = $(real(poly)) $(imag(poly))")

    gauge_action = GaugeAction(U)
    plaqloop = make_loops_fromname("plaquette")
    append!(plaqloop, plaqloop')
    β = β / 2
    push!(gauge_action, β, plaqloop)

    p = initialize_TA_Gaugefields(U) #This is a traceless-antihermitian gauge fields. This has NC^2-1 real coefficients. 
    Uold = similar(U)
    substitute_U!(Uold, U)
    MDsteps = 100
    numaccepted = 0

    numtrj = 10
    for itrj = 1:numtrj
        t = @timed begin
            accepted = MDstep!(gauge_action, U, p, MDsteps, Dim, Uold, temp1, temp2)
        end
        if get_myrank(U) == 0
            println("elapsed time for MDsteps: $(t.time) [s]")
        end
        numaccepted += ifelse(accepted, 1, 0)
        if itrj % 10 == 0
            @time plaq_t = calculate_Plaquette(U, temp1, temp2) * factor
            println("$itrj plaq_t = $plaq_t")
            poly = calculate_Polyakov_loop(U, temp1, temp2)
            println("$itrj polyakov loop = $(real(poly)) $(imag(poly))")
            println("acceptance ratio ", numaccepted / itrj)
        end
    end
    return plaq_t, numaccepted / numtrj
end

function main()
    β = 5.7
    NX = 8
    NY = 8
    NZ = 8
    NT = 8
    NC = 3
    HMC_test_4D(NX, NY, NZ, NT, NC, β)
end
main()

\end{lstlisting}

\subsection{HMC with MPI}\label{sec HMC with MPI}
Here, we show the HMC with MPI. the REPL and Jupyter notebook can not be used when one wants to use MPI. At first, in Julia REPL in the package mode,
\begin{lstlisting}[language=JuliaLocal,style=julia]
add MPIPreferences
\end{lstlisting}
and
\begin{lstlisting}[language=JuliaLocal,style=julia]
using MPIPreferences
MPIPreferences.use_system_binary()
\end{lstlisting}
With the use of 
\begin{lstlisting}[language=JuliaLocal,style=julia]
add MPI
\end{lstlisting}
we can use MPI in Julia.
We show the sample code: 
\begin{lstlisting}[language=JuliaLocal,style=julia]

using Random
using Gaugefields
using LinearAlgebra
using MPI

if length(ARGS) < 5
    error("USAGE: ","""
    mpiexecjl -np 2 exe.jl 1 1 1 2 true
    """)
end
const pes = Tuple(parse.(Int64,ARGS[1:4]))
const mpi = parse(Bool,ARGS[5])

function calc_action(gauge_action,U,p)
    NC = U[1].NC
    Sg = -evaluate_GaugeAction(gauge_action,U)/NC #evaluate_Gauge_action(gauge_action,U) = tr(evaluate_Gaugeaction_untraced(gauge_action,U))
    Sp = p*p/2
    S = Sp + Sg
    return real(S)
end

function MDstep!(gauge_action,U,p,MDsteps,Dim,Uold,temp1,temp2)
    Δτ = 1.0/MDsteps
    gauss_distribution!(p)
    Sold = calc_action(gauge_action,U,p)
    substitute_U!(Uold,U)

    for itrj=1:MDsteps
        U_update!(U,p,0.5,Δτ,Dim,gauge_action)

        P_update!(U,p,1.0,Δτ,Dim,gauge_action,temp1,temp2)

        U_update!(U,p,0.5,Δτ,Dim,gauge_action)
    end
    Snew = calc_action(gauge_action,U,p)
    if get_myrank(U) == 0
        println("Sold = $Sold, Snew = $Snew")
        println("Snew - Sold = $(Snew-Sold)")
    end
    ratio = min(1,exp(-Snew+Sold))
    r = rand()
    if mpi
        r = MPI.bcast(r, 0, MPI.COMM_WORLD)
    end
    #ratio = min(1,exp(Snew-Sold))
    if r > ratio
        substitute_U!(U,Uold)
        return false
    else
        return true
    end
end

function U_update!(U,p,ϵ,Δτ,Dim,gauge_action)
    temps = get_temporary_gaugefields(gauge_action)
    temp1 = temps[1]
    temp2 = temps[2]
    expU = temps[3]
    W = temps[4]

    for μ=1:Dim
        exptU!(expU,ϵ*Δτ,p[μ],[temp1,temp2])
        mul!(W,expU,U[μ])
        substitute_U!(U[μ],W)
        
    end
end

function P_update!(U,p,ϵ,Δτ,Dim,gauge_action,temp1,temp2) # p -> p +factor*U*dSdUμ
    NC = U[1].NC
    temp = temp1
    dSdUμ = temp2
    factor =  -ϵ*Δτ/(NC)

    for μ=1:Dim
        calc_dSdUμ!(dSdUμ,gauge_action,μ,U)
        mul!(temp,U[μ],dSdUμ) # U*dSdUμ
        Traceless_antihermitian_add!(p[μ],factor,temp)
    end
end


function HMC_test_4D(NX,NY,NZ,NT,NC,β)
    Dim = 4
    Nwing = 0
    Random.seed!(123)
    if mpi
        PEs = pes#(1,1,1,2)
        U = Initialize_Gaugefields(NC,Nwing,NX,NY,NZ,NT,condition = "hot",mpi=true,PEs = PEs,mpiinit = false) 
    else
        U = Initialize_Gaugefields(NC,Nwing,NX,NY,NZ,NT,condition = "hot")
    end

    if get_myrank(U) == 0
        println(typeof(U))
    end

    temp1 = similar(U[1])
    temp2 = similar(U[1])

    if Dim == 4
        comb = 6 #4*3/2
    elseif Dim == 3
        comb = 3
    elseif Dim == 2
        comb = 1
    else
        error("dimension $Dim is not supported")
    end
    
    factor = 1/(comb*U[1].NV*U[1].NC)

    @time plaq_t = calculate_Plaquette(U,temp1,temp2)*factor
    if get_myrank(U) == 0
        println("0 plaq_t = $plaq_t")
    end
    poly = calculate_Polyakov_loop(U,temp1,temp2) 
    if get_myrank(U) == 0
        println("0 polyakov loop = $(real(poly)) $(imag(poly))")
    end

    gauge_action = GaugeAction(U)
    plaqloop = make_loops_fromname("plaquette")
    append!(plaqloop,plaqloop')
    β = β/2
    push!(gauge_action,β,plaqloop)

    p = initialize_TA_Gaugefields(U) #This is a traceless-antihermitian gauge fields. This has NC^2-1 real coefficients. 
    Uold = similar(U)
    substitute_U!(Uold,U)
    MDsteps = 100
    temp1 = similar(U[1])
    temp2 = similar(U[1])
    comb = 6
    factor = 1/(comb*U[1].NV*U[1].NC)
    numaccepted = 0

    numtrj = 100
    for itrj = 1:numtrj
        t = @timed begin
            accepted = MDstep!(gauge_action,U,p,MDsteps,Dim,Uold,temp1,temp2)
        end
        if get_myrank(U) == 0
            println("elapsed time for MDsteps: $(t.time) [s]")
        end
        numaccepted += ifelse(accepted,1,0)

        if itrj % 10 == 0
            plaq_t = calculate_Plaquette(U,temp1,temp2)*factor
            if get_myrank(U) == 0
                println("$itrj plaq_t = $plaq_t")
            end
            poly = calculate_Polyakov_loop(U,temp1,temp2) 
            if get_myrank(U) == 0
                println("$itrj polyakov loop = $(real(poly)) $(imag(poly))")
                println("acceptance ratio ",numaccepted/itrj)
            end
        end
    end
    return plaq_t,numaccepted/numtrj

end

function main()
    β = 5.7
    NX = 8
    NY = 8
    NZ = 8
    NT = 8
    NC = 3
    HMC_test_4D(NX,NY,NZ,NT,NC,β)
end
main()
\end{lstlisting}
The command is like:
\begin{lstlisting}[language=JuliaLocal,style=julia]
mpiexecjl -np 2 julia mpi_sample.jl 1 1 1 2 true
\end{lstlisting}
We can also use MPI in LatticeDiracOperators.jl.  

\section{LatticeDiracOperators.jl}
LatticeDiracOperators.jl handles fermions on a lattice. 
This package have the following functionalities:
\begin{itemize}
    \item Constructing actions and its derivative for staggered fermion with 1-8 tastes with the use of the rational HMC
    \item Constructing actions and its derivative for Wilson fermion
    \item Constructing actions and its derivative for Standard Domainwall fermion
    \item Hybrid Monte Carlo method with fermions
\end{itemize}
With the use of the Gaugefields.jl, we can also do the HMC with stout smearing. 
This package can be regarded as the additional package of the Gaugefields.jl to treat with lattice fermions.
\subsubsection{Definition of pseudo-fermion fields}
The pseudo-fermion fields can be defined as 
\begin{lstlisting}[language=JuliaLocal,style=julia]
using Gaugefields
using LatticeDiracOperators

NX = 4
NY = 4
NZ = 4
NT = 4
Nwing = 0
Dim = 4
NC = 3

U = Initialize_4DGaugefields(NC,Nwing,NX,NY,NZ,NT,condition = "cold")
x = Initialize_pseudofermion_fields(U[1],"Wilson")
\end{lstlisting}
Here, x is a pseudo fermion fields for Wilson Dirac operator. 
The element of x is x[ic,ix,iy,iz,it,ialpha]. 
ic is an index of the color. ialpha is the internal degree of the gamma matrix. 
The staggered fermions can be defined as 
\begin{lstlisting}[language=JuliaLocal,style=julia]
x = Initialize_pseudofermion_fields(U[1],"staggered")
\end{lstlisting}
If one wants to obtain the Gaussian distributed pseudo-fermions, the code is written as 
\begin{lstlisting}[language=JuliaLocal,style=julia]
gauss_distribution_fermion!(x)
\end{lstlisting}

\subsection{Definition of Dirac operators}
The Dirac operators are important basic parts in lattice QCD simulations. 
The Wilson Dirac operator can be defined as
\begin{lstlisting}[language=JuliaLocal,style=julia]
params = Dict()
params["Dirac_operator"] = "Wilson"
params["κ"] = 0.141139
params["eps_CG"] = 1.0e-8
params["verbose_level"] = 2
D = Dirac_operator(U,x,params)
\end{lstlisting}
We can treat the Dirac operator as a matrix. 
Thus, we can apply the Dirac operator to the pseudo-fermion fields as follows.
\begin{lstlisting}[language=JuliaLocal,style=julia]
using LinearAlgebra
y = similar(x)
mul!(y,D,x)
\end{lstlisting}
And we can solve the equation $Dx = b$: 
\begin{lstlisting}[language=JuliaLocal,style=julia]
solve_DinvX!(y,D,x)
\end{lstlisting}
The convergence property can be seen by setting "\texttt{verbose\_level}" flag: 
\begin{lstlisting}[language=JuliaLocal,style=julia]
params["verbose_level"] = 3
D = Dirac_operator(U,x,params)
gauss_distribution_fermion!(x)
solve_DinvX!(y,D,x)
println(y[1,1,1,1,1,1])
\end{lstlisting}
The adjoint of the Dirac operator $D^{\dagger}$ and $D^{\dagger} D$ operator can be defined as 
\begin{lstlisting}[language=JuliaLocal,style=julia]
gauss_distribution_fermion!(x)
solve_DinvX!(y,D',x)
println(y[1,1,1,1,1,1])
DdagD = DdagD_operator(U,x,params)
gauss_distribution_fermion!(x)
solve_DinvX!(y,DdagD,x) 
println(y[1,1,1,1,1,1])
\end{lstlisting}
We can similarly define the Dirac operator for the staggered fermion as follows:
\begin{lstlisting}[language=JuliaLocal,style=julia]
x = Initialize_pseudofermion_fields(U[1],"staggered")
gauss_distribution_fermion!(x)
params = Dict()
params["Dirac_operator"] = "staggered"
params["mass"] = 0.1
params["eps_CG"] = 1.0e-8
params["verbose_level"] = 2
D = Dirac_operator(U,x,params)
y = similar(x)
mul!(y,D,x)
println(y[1,1,1,1,1,1])
solve_DinvX!(y,D,x)
println(y[1,1,1,1,1,1])
\end{lstlisting}
The "tastes" of the staggered fermion is defined in the action.

\subsection{Definition of fermion actions}
With the use of the LatticeDiracOperators.jl, we can define actions for pseudo-fermions. 
The sample codes are written as 
\begin{lstlisting}[language=JuliaLocal,style=julia]
NX = 4
NY = 4
NZ = 4
NT = 4
Nwing = 0
Dim = 4
NC = 3

U = Initialize_4DGaugefields(NC, Nwing, NX, NY, NZ, NT, condition="cold")
x = Initialize_pseudofermion_fields(U[1], "Wilson")
gauss_distribution_fermion!(x)

params = Dict()
params["Dirac_operator"] = "Wilson"
params["κ"] = 0.141139
params["eps_CG"] = 1.0e-8
params["verbose_level"] = 2

D = Dirac_operator(U, x, params)

parameters_action = Dict()
fermi_action = FermiAction(D, parameters_action)
\end{lstlisting}
Then, the fermion action with given pseudo-fermion fields is evaluated as
\begin{lstlisting}[language=JuliaLocal,style=julia]
Sfnew = evaluate_FermiAction(fermi_action,U,x)
println(Sfnew)
\end{lstlisting}
We can also calculate the derivative of the fermion action $dSf/dU$ as 
\begin{lstlisting}[language=JuliaLocal,style=julia]
UdSfdUμ = calc_UdSfdU(fermi_action,U,x)
\end{lstlisting}
The function \texttt{calc\_UdSfdU} calculates the $U dSf/dU$.
We can also use \texttt{calc\_UdSfdU!(UdSfdU$\mu$,fermi\_action,U,x)}.

In the case of the staggered fermion, we can choose "taste". 
The action is defined as 
\begin{lstlisting}[language=JuliaLocal,style=julia]
x = Initialize_pseudofermion_fields(U[1],"staggered")
gauss_distribution_fermion!(x)
params = Dict()
params["Dirac_operator"] = "staggered"
params["mass"] = 0.1
params["eps_CG"] = 1.0e-8
params["verbose_level"] = 2
D = Dirac_operator(U,x,params)

Nf = 2

println("Nf = $Nf")
parameters_action = Dict()
parameters_action["Nf"] = Nf
fermi_action = FermiAction(D,parameters_action)
Sfnew = evaluate_FermiAction(fermi_action,U,x)
println(Sfnew)
UdSfdUμ = calc_UdSfdU(fermi_action,U,x)
\end{lstlisting}
This package uses the RHMC techniques to consider the tastes.

\subsection{Hybrid Monte Carlo}
We show a sample code for the Hybrid Monte Carlo method with pseudo-fermion fields. 
The codes are written as 
\begin{lstlisting}[language=JuliaLocal,style=julia]
using Gaugefields
using LatticeDiracOperators
using LinearAlgebra
using InteractiveUtils
using Random

function MDtest!(gauge_action,U,Dim,fermi_action,η,ξ)
    p = initialize_TA_Gaugefields(U) #This is a traceless-antihermitian gauge fields. This has NC^2-1 real coefficients. 
    Uold = similar(U)
    substitute_U!(Uold,U)
    MDsteps = 10
    temp1 = similar(U[1])
    temp2 = similar(U[1])
    comb = 6
    factor = 1/(comb*U[1].NV*U[1].NC)
    numaccepted = 0
    Random.seed!(123)

    numtrj = 10
    for itrj = 1:numtrj
        @time accepted = MDstep!(gauge_action,U,p,MDsteps,Dim,Uold,fermi_action,η,ξ)
        numaccepted += ifelse(accepted,1,0)

        plaq_t = calculate_Plaquette(U,temp1,temp2)*factor
        println("$itrj plaq_t = $plaq_t")
        println("acceptance ratio ",numaccepted/itrj)
    end
end

function calc_action(gauge_action,U,p)
    NC = U[1].NC
    Sg = -evaluate_GaugeAction(gauge_action,U)/NC 
    Sp = p*p/2
    S = Sp + Sg
    return real(S)
end


function MDstep!(gauge_action,U,p,MDsteps,Dim,Uold,fermi_action,η,ξ)
    Δτ = 1/MDsteps
    NC,_,NN... = size(U[1])
    
    gauss_distribution!(p)
    
    substitute_U!(Uold,U)
    gauss_sampling_in_action!(ξ,U,fermi_action)
    sample_pseudofermions!(η,U,fermi_action,ξ)
    Sfold = real(dot(ξ,ξ))
    println("Sfold = $Sfold")

    Sold = calc_action(gauge_action,U,p) + Sfold
    println("Sold = ",Sold)

    for itrj=1:MDsteps
        U_update!(U,p,0.5,Δτ,Dim,gauge_action)

        P_update!(U,p,1.0,Δτ,Dim,gauge_action)
        P_update_fermion!(U,p,1.0,Δτ,Dim,gauge_action,fermi_action,η)

        U_update!(U,p,0.5,Δτ,Dim,gauge_action)
    end
    Sfnew = evaluate_FermiAction(fermi_action,U,η)
    println("Sfnew = $Sfnew")
    Snew = calc_action(gauge_action,U,p) + Sfnew

    println("Sold = $Sold, Snew = $Snew")
    println("Snew - Sold = $(Snew-Sold)")
    accept = exp(Sold - Snew) >= rand()
    if accept != true #rand() > ratio
        substitute_U!(U,Uold)
        return false
    else
        return true
    end
end

function U_update!(U,p,ϵ,Δτ,Dim,gauge_action)
    temps = get_temporary_gaugefields(gauge_action)
    temp1 = temps[1]
    temp2 = temps[2]
    expU = temps[3]
    W = temps[4]

    for μ=1:Dim
        exptU!(expU,ϵ*Δτ,p[μ],[temp1,temp2])
        mul!(W,expU,U[μ])
        substitute_U!(U[μ],W)
        
    end
end

function P_update!(U,p,ϵ,Δτ,Dim,gauge_action) # p -> p +factor*U*dSdUμ
    NC = U[1].NC
    temps = get_temporary_gaugefields(gauge_action)
    dSdUμ = temps[end]
    factor =  -ϵ*Δτ/(NC)

    for μ=1:Dim
        calc_dSdUμ!(dSdUμ,gauge_action,μ,U)
        mul!(temps[1],U[μ],dSdUμ) # U*dSdUμ
        Traceless_antihermitian_add!(p[μ],factor,temps[1])
    end
end

function P_update_fermion!(U,p,ϵ,Δτ,Dim,gauge_action,fermi_action,η)
    temps = get_temporary_gaugefields(gauge_action)
    UdSfdUμ = temps[1:Dim]
    factor =  -ϵ*Δτ
    calc_UdSfdU!(UdSfdUμ,fermi_action,U,η)

    for μ=1:Dim
        Traceless_antihermitian_add!(p[μ],factor,UdSfdUμ[μ])
    end
end

function test1()
    NX = 4
    NY = 4
    NZ = 4
    NT = 4
    Nwing = 0
    Dim = 4
    NC = 3
    U = Initialize_4DGaugefields(NC,Nwing,NX,NY,NZ,NT,condition = "cold")
    gauge_action = GaugeAction(U)
    plaqloop = make_loops_fromname("plaquette")
    append!(plaqloop,plaqloop')
    β = 5.5/2
    push!(gauge_action,β,plaqloop)
    show(gauge_action)
    x = Initialize_pseudofermion_fields(U[1],"Wilson")

    params = Dict()
    params["Dirac_operator"] = "Wilson"
    params["κ"] = 0.141139
    params["eps_CG"] = 1.0e-8
    params["verbose_level"] = 2
    D = Dirac_operator(U,x,params)
    parameters_action = Dict()
    fermi_action = FermiAction(D,parameters_action)
    y = similar(x)
    MDtest!(gauge_action,U,Dim,fermi_action,x,y)
end
test1()
\end{lstlisting}
We can easily switch the Wilson fermion to the staggered fermions. 

\section{QCDMeasurements.jl}
This package has following functionalities
\begin{itemize}
    \item Plaquette measurement.
    \item Polyakov loop measurement.
\item Pion correlator measurement.
\item Chiral condensate measurement.
\item Topological charge measurement.
\item Energy density measurement.
\item Wilson loop measurement
\end{itemize}
\subsection{Sample}

To measure variables, one have to make an instance 
\begin{lstlisting}[language=JuliaLocal,style=julia]
m_plaq = Plaquette_measurement(U)
\end{lstlisting}
By using this \texttt{m\_plaq}, one can measure and get the plaquette: 
\begin{lstlisting}[language=JuliaLocal,style=julia]
plaq = get_value(measure(m_plaq,U))
\end{lstlisting}

The example is shown as follows. 
\begin{lstlisting}[language=JuliaLocal,style=julia]
using QCDMeasurements
using Gaugefields
function test()
    println("SU3test")
    NX = 4
    NY = 4
    NZ = 4
    NT = 4
    Nwing = 0
    Dim = 4
    NC = 3

    U = Initialize_4DGaugefields(NC,Nwing,NX,NY,NZ,NT,condition = "cold")
    filename = "testconf.txt"
    L = [NX,NY,NZ,NT]
    load_BridgeText!(filename,U,L,NC)
    m_plaq = Plaquette_measurement(U)
    m_poly = Polyakov_measurement(U)

    plaq = get_value(measure(m_plaq,U))
    poly = get_value(measure(m_poly,U))
    println("plaq: $plaq")
    println("poly: $poly")

    m_energy = Energy_density_measurement(U)
    m_topo = Topological_charge_measurement(U)
    energy = get_value(measure(m_energy,U))
    topo = get_value(measure(m_topo,U))
    println("energy: $energy")
    println("topo: $topo")

    m_wilson = Wilson_loop_measurement(U,printvalues=true)
    wilsonloop = get_value(measure(m_wilson,U))
    println("wilson loop: ",wilsonloop)

    m_pion = Pion_correlator_measurement(U)
    m_pion_Staggered = Pion_correlator_measurement(U,fermiontype = "Staggered")
    m_pion_Wilson = Pion_correlator_measurement(U,fermiontype = "Wilson")
    pion = get_value(measure(m_pion,U))
    pion_s = get_value(measure(m_pion_Staggered,U))
    pion_w = get_value(measure(m_pion_Wilson,U))

    println("pion: $pion")
    println("pion correlator with Staggered fermion: $pion_s")
    println("pion correlator with  Wilson fermion: $pion_w")

    m_chiral_Staggered = Chiral_condensate_measurement(U,fermiontype = "Staggered")
    m_chiral_Wilson = Chiral_condensate_measurement(U,fermiontype = "Wilson")
    chiral_s = get_value(measure(m_chiral_Staggered,U))
    chiral_w = get_value(measure(m_chiral_Wilson,U))

    println("Chiral condensate with Staggered fermion: $chiral_s")
    println("Chiral condensatewith  Wilson fermion: $chiral_w")


    TC_methods = ["plaquette","clover"]
    m_topo = Topological_charge_measurement(U,TC_methods = TC_methods)
    g = Gradientflow(U)
    for itrj=1:100
        flow!(U,g)
        @time plaq_t = get_value(measure(m_plaq,U))
        @time poly = get_value(measure(m_poly,U))
        println("$itrj plaq_t = $plaq_t")
        println("$itrj polyakov loop = $(real(poly)) $(imag(poly))")

        @time topo = get_value(measure(m_topo,U))
        print("$itrj topological charge: ")
        for (key,value) in topo
            print("$key $value \t")
        end
        println("\t")
    end

end
test()
\end{lstlisting}
One can also use the dictionary type:
\begin{lstlisting}[language=JuliaLocal,style=julia]
using QCDMeasurements
using Gaugefields
function SU3test()
    println("SU3test")
    NX = 4
    NY = 4
    NZ = 4
    NT = 4
    Nwing = 0
    Dim = 4
    NC = 3

    U = Initialize_4DGaugefields(NC,Nwing,NX,NY,NZ,NT,condition = "cold")
    filename = "testconf.txt"
    L = [NX,NY,NZ,NT]
    load_BridgeText!(filename,U,L,NC)

    method = Dict()
    methodname = "Eigenvalue"
    method["methodname"] = methodname
    method["fermiontype"] = "Wilson"
    κ = 0.141139
    method["hop"] =  κ
    method["nev"] = 1 #number of eigenvalues
    m = prepare_measurement_from_dict(U,method)
    value,vectors = get_value(measure(m,U)) #eigenvalues and eigenvectors
    println("$methodname $value")
    

    method = Dict()
    methodname = "Pion_correlator"
    method["methodname"] = methodname
    method["fermiontype"] = "Staggered"
    method["mass"] = 1
    method["Nf"] = 4
    m = prepare_measurement_from_dict(U,method)
    value = get_value(measure(m,U))
    println("$methodname $value")

    method = Dict()
    methodname = "Pion_correlator"
    method["methodname"] = methodname
    method["fermiontype"] = "Wilson"
    method["hop"] = 1
    m = prepare_measurement_from_dict(U,method)
    value = get_value(measure(m,U))
    println("$methodname $value")


    methodsname = ["Plaquette","Polyakov_loop","Topological_charge","Chiral_condensate",
            "Pion_correlator","Energy_density","Wilson_loop","Eigenvalue"]
    method = Dict()
    for methodname in methodsname
        method["methodname"] = methodname
        m = prepare_measurement_from_dict(U,method)
        value = get_value(measure(m,U))
        if methodname == "Eigenvalue"
            println("$methodname $(value[1])")
        else
            println("$methodname $(value)")
        end
    end

end
SU3test()
\end{lstlisting}

\section{Consistency check} \label{sec:Consistency check}
We compared results of LatticeQCD.jl to the following papers and codes
\begin{itemize}
    \item $N_f=4$ SU(3) staggered HMC \cite{Gavai1989-qq} 
    \item Quenched SU(2) improved thermodynamics \cite{Giudice2017-nj}
    \item RHMC \cite{Tomiya2018-iy} 
    \item HMC for Wilson and Clover Wilson fermions \href{https://github.com/tsuchim/Lattice-Tool-Kit}{Lattice Tool Kit}
    \item Pion correlator with the Wilson-Dirac operator \cite{Butler1994-ib} 
    \item Pion correlator with the staggered Dirac operator \cite{Bowler1988-tw}  
\end{itemize}

\end{document}